# Impact of chain fragmentation on charge transfer scenario and two-plateaus-like behavior of $T_c(x)$ in $YBa_2Cu_3O_{6+x}$

V. M. Matic and N. Dj. Lazarov

*Laboratory of Theoretical Physics,
Institute of Nuclear Sciences "Vinca", 11001 Belgrade, Serbia*

**Abstract**

A model of charge transfer mechanism from CuO chains to $CuO_2$ planes has been proposed to account for doping of the planes, assuming that only chains containing more than three oxygen atoms can contribute to hole transfer. Only chains with $l \geq 4$ are assumed to have transferred a certain fraction, approximately 40%, of the holes created by oxygen atoms added to the chain beyond the first three oxygen atoms. Using the so obtained $x$ dependence of doping, $p(x)$, at constant (room) temperature and utilizing empirical parabolic phase relation $T_c(p)$ ($T_c(p)=T_{c,max}[1-82.6(p(x)-0.16)^2]$), the $T_c$ versus $x$ dependence is obtained to have two clearly distinguished plateaus at 60K and 90K, remarkably fitting to experimental $T_c(x)$. The effect of statistics of CuO chain fragmentation has been included by applying cluster variation method to two dimensional asymmetric next nearest neighbor Ising model that is employed to describe oxygen-chain ordering in basal planes. The obtained results indicate that plateaus, coinciding with $T_c'(x)=0$, emerge either when $p'(x)=0$ ($p(x) \approx const$), in the region of OII phase formation (the 60K plateau), or when $p=0.16$, representing the optimal doping at $x \approx 0.91$ in OI phase (the 90K plateau).

YBa$_2$Cu$_3$O$_{6+x}$ superconductor occupies unique position among other high-$T_c$ cuprates because doping of CuO$_2$ layers is achieved not by chemical substitution of metal atoms that are sandwiched between the layers, but by addition of oxygen that orders to form CuO chains. The chains are known to act as a reservoir of positive charge that is subsequently transferred to the layers. For example, in an underdoped regime opening of pseudo gap in the normal state at $T^*>T_c$ [1] is associated with charge fluctuations in layers and chains, which are thought that might be due to charge density waves (CDW) [2-4], that trigger, at a certain stage, the charge transfer process from chains to planes during temperature decrease [2]. The idea that the ability of a chain to supply holes to the layers should explicitly depend on its length is well grounded on experiments with photoinduced superconductivity [5-8] and room temperature aging of YBa$_2$Cu$_3$O$_{6+x}$ samples that had previously been fast cooled from high temperatures [9,10], in which corresponding rise of $T_c$ was accompanied by reordering of oxygen in basal planes to form longer chains. On the other hand, the generally promoted idea, often used to embrace the emergence of 60K plateau in $T_c(x)$ within reasonable acceptable explanation, is founded on a belief that short chains should not be considered as efficient hole dopands [11-13]. According to this concept, there should exists a minimal chain length, that we will denote here as *critical chain length* $l_{cr}$, below which there is no hole transfer, so that only chains with $l \geq l_{cr}$, can contribute to doping. Such a point of view has also been supported by theoretical studies on band structure calculations of CuO chain fragments showing that they tend to become unstable with respect to charge transfer only when their length is increased beyond a certain threshold value [14-16]. However, a coherent agreement on what the value of $l_{cr}$ might be, as well as on how to count the contribution of long chains $l > l_{cr}$ to doping of the CuO$_2$ layers, has not been achieved so far.

In order to evaluate the number of holes transferred it is necessary to find out how many chains in the system are short ($l<l_{cr}$), and how many are long ($l \geq l_{cr}$). In other words, resolving the issue of length distribution of CuO chains $f(l)$ emerges as a required first step in obtaining $x$ dependences of both the doping $p(x)$ and critical transition temperature $T_c(x)$. Unfortunately, very little has been done over the years on elucidating the real nature of $f(l)$ although it was sometimes speculated that it might have had a maximum at $l=\langle l \rangle$ (average chain length) spreading out around it in, more or less, a Gaussian-like fashion, so that in some studies it has even been proposed that all chains have nearly the same length, $l \approx \langle l \rangle$, implicitly assuming the possible Gaussian $f(l)$ to be narrow enough [17]. However, the problem of $l$ dependence of length distribution has recently been addressed showing $f(l)$ to be maximal for shortest chains $l=1$ (isolated oxygen atoms) further decreasing monotonously with $l$ in the manner of *geometric progression* [18]

$$f_m(l) = \frac{1}{\langle l_m \rangle}\left(1 - \frac{1}{\langle l_m \rangle}\right)^{l-1} \quad , \quad m = 1, 2, \ldots \qquad (1)$$

where $\langle l_m \rangle$ denotes average chain length along nonequivalent oxygen chain site sublattices in the basal plane. The relevance of (1) for doping $p(x)$, and, consequently, two plateaus behavior of $T_c(x)$, is in that it implies charge transfer to be suppressed to a higher degree than in the case of, say, fictitious Gausian $f(l)$, for short chains are concentrated below the maximum of the distribution rather than at its edge.

Here we show that chain length distribution (1), when combined with universal phase relation on $p$ dependence of $T_c$ [19]

$$T_c(p) = T_{c,\max}\left[1 - 82.6(p(x) - 0.16)^2\right], \quad (T_{c,\max}=93K), \tag{2}$$

yields two plateaus form of $T_c(x)=T_c(p(x))$, where $x$ dependence of doping, $p(x)$, at $T=const$ (=room temperature), is determined through concentration of those holes in chains that are susceptible to transfer towards layers. The results obtained point to the conclusion that 60K plateau is connected with constant doping, $p(x)\approx const$, due to OII phase formation, while 90K plateau coincides with optimal doping level at $p=0.16$ holes per Cu site (from (1) it follows $T_c'(x)=0$ if either $p'(x)=0$, or $p=0.16$). We also show that the two main orthorombic phases, OI and OII, are responsible for emergence of two plateaus in $T_c(x)$, while other ortho phases, OIII, OIV, OV and OVIII, that have been reported in experiments [20,21], are likely to be of a minor importance for such pronounced characteristics of $T_c(x)$, and they might be of a some influence only in transition regime where $T_c$ undergoes abrupt change. To say in other words, we claim that the bare ASYNNNI model (asymmetric next-to-nearest neighbor Ising model) of de Fontaine and Wille [22] suffices to account for both plateaus only if the parameter $l_{cr}$ attains optimal value.

It is a commonly accepted opinion that in the chain (basal) plane Cu can be either +1, which is the case when it is connected to only two apical O(4) atoms (2-fold coordinated Cu), or +2, when copper ion is 3-fold coordinated (located at chain end) and 4-fold coordinated (located within the chain). When isolated oxygen occupies a site in the basal plane, it then takes two electrons from two nearest Cu ions transforming them from $Cu^{+1}$ to $Cu^{+2}$. Accordingly, isolated oxygen, representing itself a chain of length $l=1$, neither creates holes, nor it is capable of transferring them, since no local charge balance has been violated in such a process. When another oxygen is added to form $l=2$ chain there is only one electron available in nearby copper coordination so that one hole appears to be created. In this way, a chain of length $l$ is seen to have created $l-1$ holes. Here we propose, applying the concept of critical chain length, that no holes can be transferred from chains with $l<l_{cr}$, while from chains with length $l\geq l_{cr}$ only $l-(l_{cr}-1)$ holes are susceptible to transfer. We further propose that only a certain fraction of these transferable holes, say $\chi$, will really be transferred to the layers contributing to their doping, so that the number of transferred holes from chain of length $l$ is equal to $P(l)=\chi(l-l_{cr}+1)$. From Figure 1 it is obvious that the number of oxygen chain sites is equal to the number of Cu ions in the basal plane $N_{Cu}$. If $n$ denotes fraction of 3-fold coordinated Cu it follows that $(n/2)N_{Cu}$ is the total number of chains. $f(l)(n/2)N_{Cu}$ is then equal to the total number of chains with a given length $l$. Taking into account that the doping $p$ is usually defined as the number of transferred holes per Cu, it follows

$$p = \chi \frac{n}{4} \sum_{l=l_{cr}}^{\infty} (l - l_{cr} + 1) f(l). \tag{3}$$

Here, the factor 1/4 is introduced (instead of 1/2) to include the fact that in the double-layered high-$T_c$ compound $YBa_2Cu_3O_{6+x}$ an each chain plane feeds with holes two $CuO_2$ layers which lie immediately above and below it. In the case of OII phase, when oxygen chain sites ($\alpha$) split into two interlacing sublattices, that are commonly denoted by $\alpha_1$ and $\alpha_2$ and are characterized by different rates of chain breaking $n_1$ and $n_2$ ($n=\frac{1}{2}(n_1+n_2)$) and different oxygen occupations $x_1$ and $x_2$ ($x=\frac{1}{2}(x_1+x_2)$), the length distribution $f(l)$ is given by

$$f(l) = \frac{n_1 f_1(l) + n_2 f_2(l)}{n_1 + n_2}, \qquad (4)$$

where $f_1(l)$ and $f_2(l)$ are determined by (1), while average chain lengths $\langle l_1 \rangle$ and $\langle l_2 \rangle$ are equal to $\langle l_1 \rangle = 2x_1/n_1$ and $\langle l_2 \rangle = 2x_2/n_2$, respectively. Inserting (4) and (1) into (3) and making use of the well known result for the sum of geometric progression one arrives at

$$p = \frac{\chi}{4}\left[ x_1\left(\frac{2x_1 - n_1}{2x_1}\right)^{l_{cr}-1} + x_2\left(\frac{2x_2 - n_2}{2x_2}\right)^{l_{cr}-1} \right]. \qquad (5)$$

From the above expression and (2) it can be seen that the doping, and, hence, the $T_c$, depend on two parameters $\chi$ and $l_{cr}$, beside the quantities $x_1$, $x_2$, $n_1$, and $n_2$ that reflect thermodynamics of oxygen ordering. We used 5/4 point approximation of cluster variation method (CVM) applied to the ASYNNNI model to calculate these four quantities at given values of $x$ and $T$. To get a clue of what the value of parameter $\chi$ might be it is worthwhile to observe that from (5) it follows $p \to \chi/2$ as $x \to 1$ (OI phase) regardless of the value of $l_{cr}$. Although the doping is not convincingly measurable in experiments around $x \approx 0.5$ [11,19] (OII phase), it is nevertheless conveniently obtainable at higher oxygen content, at approximately $x > 0.7$, so $p(x=1) \approx 0.187$ was obtained in Reference [19]. Thus, it can be expected $\chi \approx 38\%$, or even $\chi \approx 40\%$, to be in good correlation with experiment.

We calculated $p(x)$ and $T_c(x) = T_c(p(x))$ using (5) and (2), at reduced temperature $\tau = k_B T/V_1 = 0.45 = const$, by varying parameters $l_{cr}$ and $\chi$ (slightly around $\chi = 0.40$) in order to acquire the best fitting to the experimentally obtained two-plateaus like $T_c(x)$ [23] (Figure 2a). The values of interaction constants $V_1>0$, $V_2<0$, and $V_3>0$ as obtained from linear muffin-tin orbital method [25] were used as input parameters. We have chosen the isotherm $\tau=0.45$ because we estimated it to be the best representative of room temperature, given the fact that in our CVM phase diagram (inset in Figure 1) the top of OII phase corresponds to $\tau=0.58$ while in some studies the top of OII phase was reported at $125^0C$ [20]. It can be seen from Figure 2a that calculated $T_c(x)$ for $l_{cr}=4$ remarkably fits in with standard experimental result on $T_c(x)$ of Jorgensen *et al.* [23]. Beside the two plateau features, the calculated $T_c(x)$ is as well characterized by two regions of abrupt change: one, below the OII composition $x=0.5$, accompanied by vanishing of superconductivity, $T_c=0$, at x≈0.4, and the other, at $0.65<x<0.85$. The parameter $\chi$ was obtained to be equal to 0.392, which is also in a very good correlation with results on $p(x)$, at $x \approx 1$, of Tallon *et al.* [19]. The obtained $p(x=1)=0.196$ is only slightly beyond the critical doping level $p=0.19$ that is known to coincide with disappearance of pseudogap phase, which explains why the $YBa_2Cu_3O_{6+x}$ is not manageable for reaching highly overdoped regimes. It should be noted that although the magnitudes of interactions $V_1$, $V_2$ and $V_3$ have sometimes been considered as being controversial, it is the NN interaction $V_1$ whose magnitude seems to have been the least questioned. Since there is a wide consensus that $V_1$ should rank around ≈6.5mRy [20,28], and, given the fact that it defines scaling between $T$ and $\tau$, one can accept with a great deal of confidence that the room temperature falls at some point between $\tau \approx 0.40$ and $\tau \approx 0.45$. With this in mind we also calculated $T_c(x)$ at $\tau=0.43$ which is shown in Figure 2b. The best correlation with the experiment is achieved here for $l_{cr}$=5 yielding the 60K plateau section to be more pronounced than in Figure 2a, but calculated values fall below experimental $T_c(x)$ in the transition region where $T_c$ starts its intense growth towards the maximum at optimal doping ($x_{opt}$≈0.91). We therefore take $l_{cr}$=4 to be our best estimate for $l_{cr}$. The χ parameter was found to have basically the same value, χ=0.388, as in Figure 2a. It is interesting to point out that when χ decreases the 60K plateau moves downwards, towards lower temperatures, while $x_{opt}$ shifts towards higher oxygen concentrations (beyond $x$=0.91).

In summary, we have shown it is the formation of OII phase that is responsible for 60K plateau. In addition, we have also obtained that the 60K plateau extends smoothly beyond the OII phase region into the region of OI phase before $T_c$ turns to rise abruptly. The 60K plateau is clearly due to nearly constant doping $p(x)$≈$const$ on the oxygen rich side of OII stoichiometry $x$=0.5. We therefore claim that such a behavior of $p(x)$ would have been obtained in Figure 4 of Reference [19] if $p$ were to be measured in the OII regime, around $x$≈0.5. This follows from $T_c'(x)$=0, that one expects to find at the plateaus, implying either $p'(x)$=0, or $p$=0.16 (equation (2)). The former solution corresponds to 60K plateau, while the later is connected with optimal doping at $x$≈0.91 and 90K plateau. It should be mentioned that experimental $T_c(x)$ of Jorgensen *et al* [23] was obtained long ago, almost immediately after the discovery of high-$T_c$ superconductivity, so that the finer structure of the 90K plateau, connected with maximum of $T_c$ at optimal doping $p$=0.16 (at $x_{opt}$≈0.91) was not seen in their $T_c(x)$; we nevertheless use here the $T_c(x)$ from Reference [23] for comparison for it has been frequently cited since and therefore can be regarded as being one of the "classical results" on the two-plateaus phenomenon. The most recent data on well prepared single crystal $YBa_2C_3O_{6+x}$ samples revealed the 60K "quasi-plateau" not so ideally horizontal; instead, it seems a bit inclined [11], i.e. in a similar fashion as we have obtained in Figure 2a, for $l_{cr}$=4. It is also worth mentioning that almost linear variation of $p(x)$ of related compound $YBa_{2-b}(Ca_b)Cu_3O_{6+x}$ over the region around $x$≈0.5 (also shown in Figure 4 of Reference [19]) can be successfully accounted for by the model in which the doping originating from introduced Ca and from the chains (5) are taken into consideration independently [24].

We are not of opinion that other-than OI and OII orthorombic phases would substantially violate obtained $T_c(x)$, although these phases can be included by addition of further than *NNN* repulsive Coulomb O-O interactions along the *a* axes [21]. The diffuse nature of their superstructure reflections points towards conclusion that these phases probably appear only as small patches that are embedded within large domains of the main phases OI and OII. Although the importance of other chain-like structures cannot be neglected, particularly in light of the recent findings that charge fluctuations in $CuO_2$ layers (charge stripes) use oxygen superstructures OIV and OV as spatial template [26,27], it is to be emphasized that no one of these structures except OII have been reported at x<0.62 [20,21] (our calculations show that, at τ=0.45, the OI-to-OII second-order phase transition occurs for $x$=0.63). On the other hand, the correlation length of OII phase has been found to extend over up to 70 unit cells along *a* axes in well prepared single-crystal samples [12,13].

It should be noted that our extensive Monte Carlo and CVM based analysis (not shown here) have demonstrated that the $l_{cr}$ concept and two plateaus behavior of $T_c(x)$ are intrinsically built-in within the ASYNNNI model scenario of basal plane O-ordering in the way that, at any constant temperature below the top of OII phase, there can always be uniquely determined the optimal $l_{cr}(T)$ that yields the two plateaus in $T_c(x)$, as shown in Figure 2. In order to get a notion of how the shape of $T_c(x)$ changes with $l_{cr}$ we show in Figure 3 the calculated $T_c(x)$ values at $\tau=0.45$ for $l_{cr}=3$, 5 and 6. We have found this to be a general trend at all $T=const$: for $l_{cr}$ less than optimal $l_{cr}(T)$ the 60K plateau disappears, while for $l_{cr}>l_{cr}(T)$ it transforms into a peak at $x\approx0.5$ accompanied with a minimum at oxygen rich side.

## Acknowledgement

This work has been funded by the Ministry of Science and Technology of the Republic of Serbia through the Project 141014.

## Figure Captions

**Figure 1.** Schematic description of basal plane of $YBa_2Cu_3O_{6+x}$. Open circles ($\alpha$) and squares ($\beta$) denote oxygen sites, small black circles stand for Cu ions. The $\beta$ sites are practically unoccupied for all $x$ because of strong *NN* repulsive O-O interaction $V_1>0$. At the OI stoichiometry $x=1$ all $\alpha$ sites are completely occupied ($\beta$ empty), whilst at $x=0.5$ (OII stoichiometry) all even $\alpha$ columns ($\alpha_1$) are occupied and $\alpha_2$ and $\beta$ sites are empty. The attractive *NNN* O-O interaction $V_2<0$, mediated by Cu ion, and repulsive Coulomb *NNN* interactions are also shown. Inset: The ASYNNNI model phase diagram of $YBa_2Cu_3O_{6+x}$ obtained by use of CVM.

**Figure 2.** Calculated values of $T_c(x)$ shown by dot-dashed line a) at $\tau=0.45=const$ for $l_{cr}=4$, and b) at $\tau=0.42=const$ for $l_{cr}=5$. The $T_c(x)$ obtained experimentally in Reference [23] (solid line) is also shown. The solid line stood in Reference [23] as guide-to-eye line connecting experimental points.

**Figure 3.** Calculated values of $T_c(x)$ for $l_{cr}=3$ (dashed), $l_{cr}=5$ (solid) and $l_{cr}=6$ (dot-dashed) at $\tau=0.45=const$. The optimal value of $l_{cr}$ at this temperature, i.e. the one that yields $p(x)=const$ section of the 60K plateau, is equal to 4.

## References

[1] J. L. Tallon, T. Benseman, G. V. M. Williams and J. W. Loram, Physica C 415, 9 (2004).
[2] A, Suter, M. Mali, J. Roos, D. Brinkman, J. Karpinski and E. Kaldis, Phys. Rev. B 56, 5542 (1997).
[3] B. Grevin, Y. Berthier and G. Collin, Phys. Rev. Lett. 85, 1310 (2000).

[4] M. Maki, T. Nishizaki, K. Shibata and N. Kobayashi, Phys. Rev. B 65, 140511 (2002).

[5] A. Bruchhausen, S. Bahrs, K. Fleiischer, A. R. Goffi, A. Fainstein, G. Nieva, A. A. Aligia, W. Richter and C. Thomsen, Phys. Rev. B 69, 224508 (2004).

[6] S. Bahrs, J. Guimpel, A. R. Gofii, B. Maiorov, A. Fainstein, G. Nieva and C. Thompson, Phys. Rev. B 72, 144501 (2005).

[7] M. Osada, M. Kall, J. Backstrom, M. Kakihana, N. H. Andersen and L. Borjesson, Phys. Rev. B 71, 214503 (2005).

[8] H. W. Seo, Q. Y. Chen, M. N. Iliev, T. H. Johansen, N. Kolev, U. Welp, C. Wang and W. K.Chu, Phys. Rev. B 72, 052501 (2005).

[9] R. Shaked, J. D. Jorgensen, B. A. Hunter, R. L. Hitterman, A. P. Paulikas and B. W. Veal, Phys. Rev. B 51, 547 (1995).

[10] A. Knizhnik, G. M. Reisner and Y. Eckstein, J. Phys. Chem. of Solids 66, 1137 (2005).

[11] R. Liang, D. A. Bohn and W. N. Hardy, Phys. Rev. B 73, 180505 (2006).

[12] F. Yakhou, J. Y. Henry, P. Burlet, V.P. Plakhty, M. Vlasov and S. Moshkin, Physica C 333, 146 (2000).

[13] Z. Yamani, W. A. McFarlane, B. W. Statt, D. Bohn, R. Liang and W. N. Hardy, Physics C 405, 227 (2004).

[14] J. Zaanen, A. T. Paxton and O. K. Andersen, Phys. Rev. Lett. 60, 2685 (1988).

[15] P. Gawiec, D. R. Grempel, A. C. Riiser, H. Haugerud and G. Uimin, Phys. Rev. B 53, 5872 (1996).

[16] P. Gawiec, D. R. Grempel, G. Uimin and J. Zittartz, Phys. Rev. B 53, 5880 (1996).

[17] E. E. Tornau, S. Lapinskas, A. Rosengren and V. M. Matic, Phys. Rev. B 49, 15952 (1994).

[18] V. M. Matic and N. Dj. Lazarov, Physica C 443, 49 (2006).

[19] J. L. Tallon, C. Bernhard, H. Shaked, R. L. Hitterman and J. D. Jorgensen, Phys. Rev. B 51, 12911 (1995).

[20] M.v. Zimmermann, J. R. Schneider, T. Frelo, N. H. Andersen, J. Madsen, J. Kall, H. F. Poulsen, R. Liang, P. Dosanjih and W. N. Hardy, Phys. Rev. B 68, 104515 (2003).

[21] P. Manca, S. Sanna, G. Calestani, A. Migliori, S. Lapinskas and E. E. Tornau, Phys. Rev. B 63, 134512 (2001).

[22] L. T. Wille and D. de Fontaine, Phys. Rev. B 37, 2227 (1988).

[23] J. D. Jorgensen, M. A. Beno, D. G. Hinks, L. Soderholm, K. J. Volin, R. L. Hitterman, J. D. Grace, J. K. Schulle, C. U. Segre, K. Zhang and M. S. Kleefisch, Phys. Rev. B 36, 3608 (1987).

[24] V. M. Matic and N. Dj. Lazarov, to be submitted.

[25] P. A. Sterne and L. T. Wille, Physica C 162-164, 223 (1989).

[26] Z. Islam, S. K. Sinha, D. Haskel, J. C. Lang, G. Srajer, B. W. Veal, D. R. Haeffner and H. A. Mook, Phys. Rev. B 66, 092501 (2002).

[27] Z. Islam, X. Liu, S. K. Sinha, J. C. Lang, S. C. Moss, D. Haskel, G. Srajer, P. Wochner, D. R. Lee, D. R. Haeffner and U. Welp, Phys. Rev. Lett. 93, 157008 (2004).

[28] D. J. Liu, L. T. Einstein, P. A. Sterne and L. T. Wille, Phys. Rev. B 52, 9784 (1995).

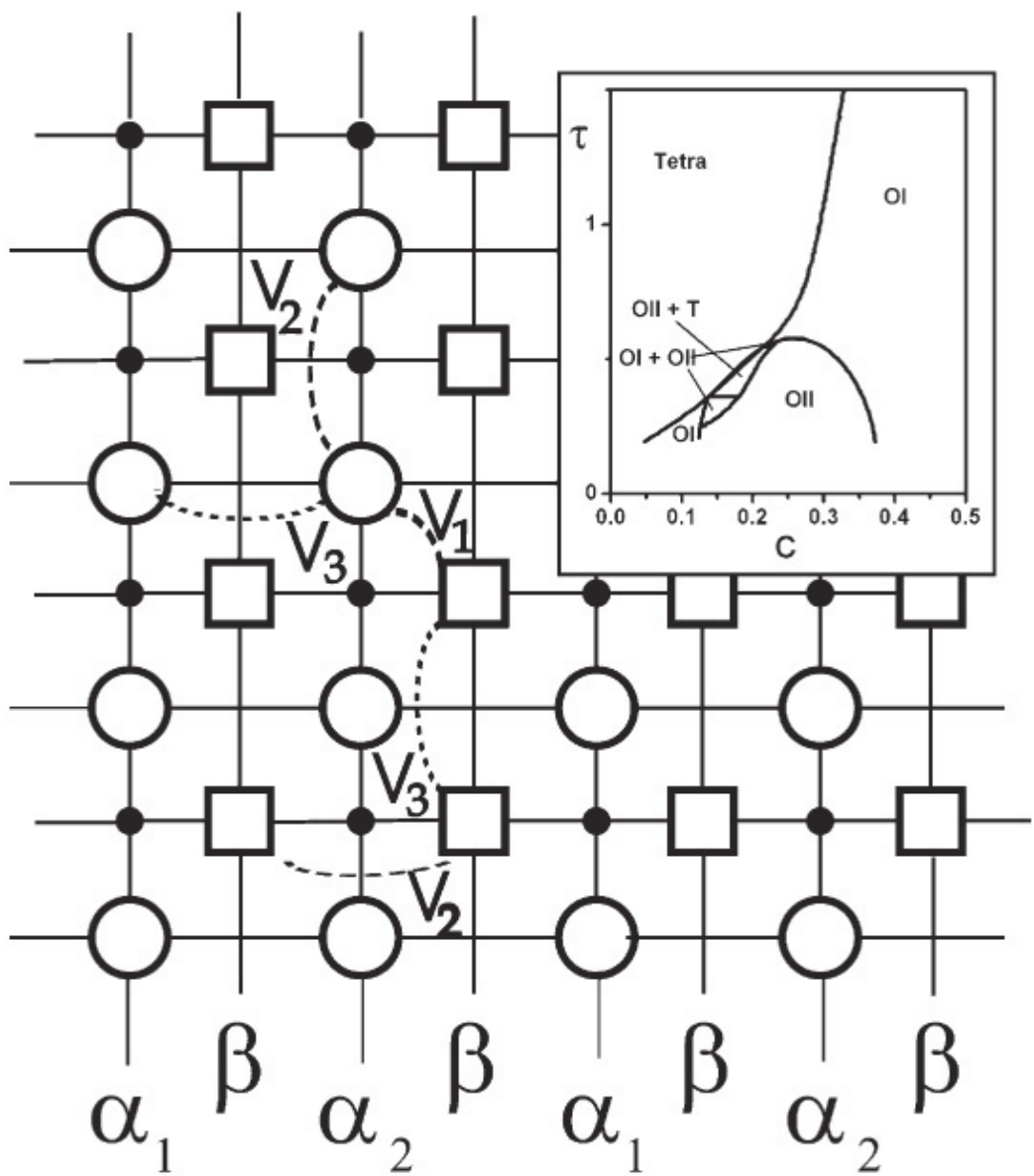

Figure 1

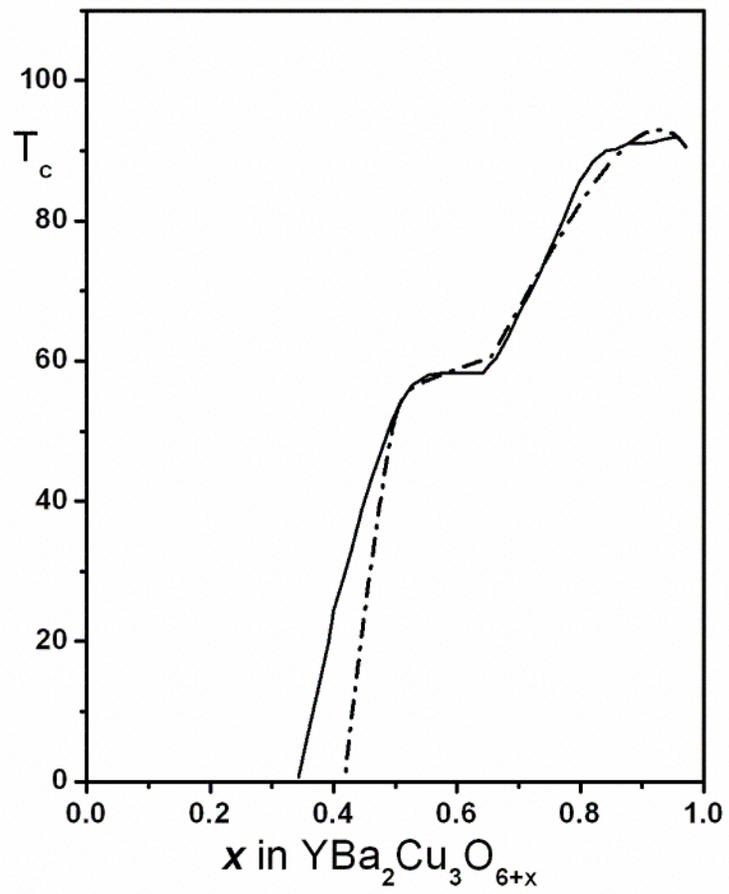

**Figure 2a**

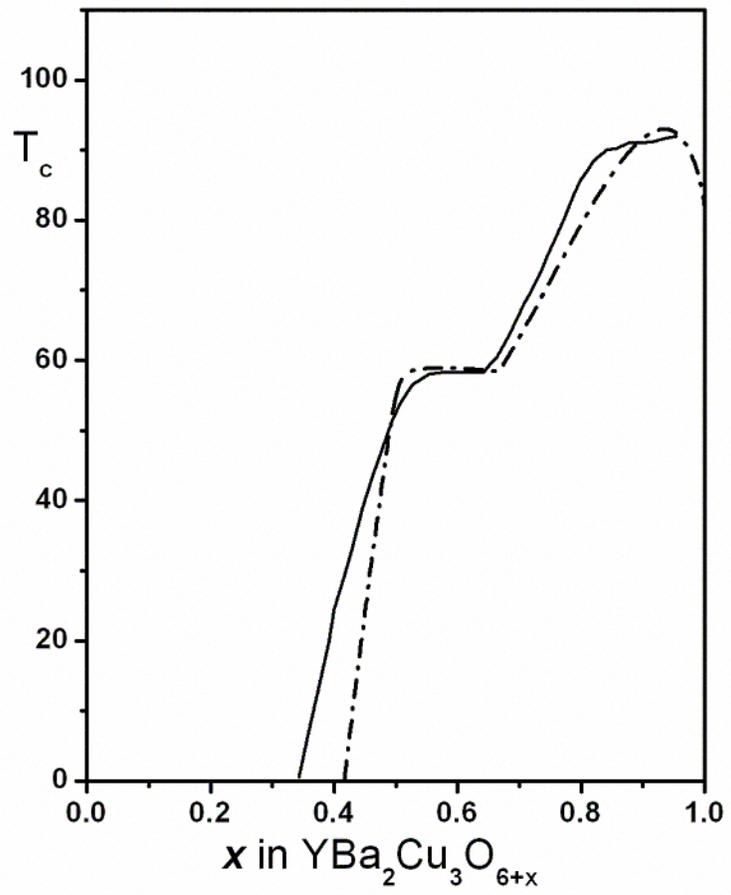

**Figure 2b**

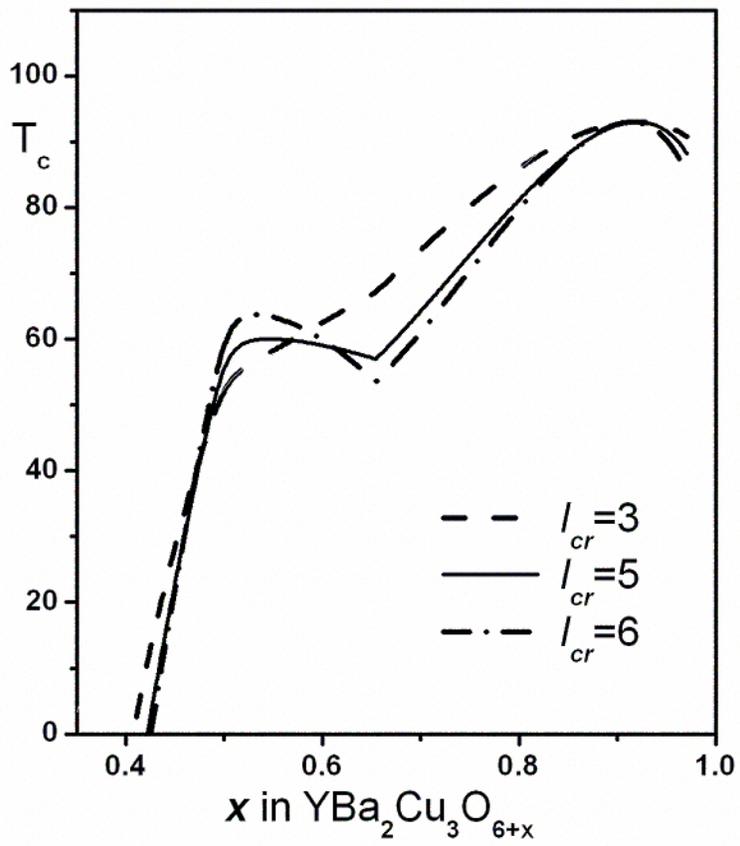

**Figure 3**